\documentclass[sigconf, nonacm]{acmart}

\usepackage{subfigure}
\AtBeginDocument{%
  \providecommand\BibTeX{{%
    \normalfont B\kern-0.5em{\scshape i\kern-0.25em b}\kern-0.8em\TeX}}}

\begin{document}

\title[ ]{A Simulation Platform for Multi-tenant Machine Learning Services on Thousands of GPUs}



\author{Ruofan Liang}
\affiliation{University of Toronto}
\email{ruofan@cs.toronto.edu}

\author{Bingsheng He}
\affiliation{National University of Singapore}
\email{hebs@comp.nus.edu.sg}

\author{Shengen Yan}
\affiliation{SenseTime}
\email{yanshengen@sensetime.com}

\author{Peng Sun}
\affiliation{SenseTime}
\email{sunpeng1@sensetime.com}

\renewcommand{\shortauthors}{} 

\begin{abstract}

Multi-tenant machine learning services have become emerging data-intensive workloads in data centers with heavy usage of GPU resources. Due to the large scale, many tuning parameters and heavy resource usage, it is usually impractical to evaluate and benchmark those machine learning services on real clusters. In this demonstration, we present AnalySIM, a cluster simulator that allows efficient design explorations for multi-tenant machine learning services. Specifically, by trace-driven cluster workload simulation, Analy\-SIM can easily test and analyze various scheduling policies in a number of performance metrics such as GPU resource utilization. AnalySIM simulates the cluster computational resource based on both physical topology and logical partition. The tool has been used in SenseTime to understand the impact of different scheduling policies with the trace from a real production cluster of over 1000 GPUs. We find that preemption and migration are able to significantly reduce average job completion time and mitigate the resource fragmentation problem. 
\end{abstract}


\newcommand{\HBS}[1]{{\textcolor{black}{ #1}}}

\maketitle

\section{Introduction}
Recent years have seen the fast development of AI and machine learning (ML) in many fields including computer vision, natural language processing, recommendation system and self-learning systems. As such, it is very common to see that ML models are used to provide personalized and intelligent services for customers. Thus, machine learning services have become emerging data-intensive workloads in data centers. Unlike developing traditional services,
ML services usually have an additional model training step which relies on training data and parameter tuning. To make it easier to develop ML services, many systems (such as Ease.ml \cite{li2018ease} and Rafiki \cite{wang2018rafiki}) are proposed for various use cases like model selecting, hyper-parameter tuning and inference serving.

Many machine learning services involve deep learning training, and GPUs are heavily used in training the deep models. It is common to see that service providers are equipped with hundreds or even thousands of GPUs for offering efficient model trainings. However, GPUs are still expensive resources in data centers. According to AWS EC2 pricing
a GPU instance in the cloud costs much higher (5-10x) than that of a regular instance. Efficiently utilizing and allocating GPU resources in the cluster nowadays is another important topic in both industry and academia. 

The commonly used cluster management tools like Slurm \cite{yoo2003slurm} and YARN \cite{vavilapalli2013apache} are mainly designed for traditional big data jobs, neglecting the rigid usage of GPU resources in machine learning services and GPU sharing from multiple tenants. Besides, the existing scheduling policies adapted by these cluster management tools can also cause heavy head-of-line (HOL) blocking due to the large variation on the running time of multiple submitted ML training jobs. In this way, new shared resource management methods and related job scheduling policies should be considered. Therefore, recently we have seen many studies focusing on improving the GPU utilization of multi-tenant machine learning services in data centers~\cite{peng2018optimus,xiao2018gandiva, gu2019tiresias}.

However, such research requires system-level modification and evaluation on large-scale GPU clusters, which is inefficient, costly and sometimes even risky to production systems. Besides, the emerging learning-based resource management methods like \cite{mao2019learning} requires thousands of iterations of parameter updates to learn a usable scheduling policy, which makes it almost impossible to be learned from a real running cluster. Clearly, there needs a platform or tool to study and evaluate shared resource management for ML development in data centers. Given these situations, the comprehensive simulation of data centers and resource management policies is of great need for
testing and evaluating different ideas for shared resource management.

To address the aforementioned needs for ML service developments, we develop AnalySIM, a flexible trace-driven GPU cluster resource management simulator, to support the evaluation and development of cluster scheduling policies. With detailed job trace simulation, we can efficiently evaluate existing scheduling policies and explore potential directions for future improvement. Specifically, we develop visual analyzing tools to show the simulation results of various resource management methods (e.g. migration and reallocation) and job scheduling policies (e.g. preemptive and non-preemptive scheduling).  In the demonstration, we use AnalySIM to simulate a production Slurm cluster with over 1000 GPUs from SenseTime, and demonstrate the simulation accuracy and our findings in comparing different scheduling policies. 

\section{Related Work}
\textbf{Systems for ML services.}
In recent years, many systems for ML is proposed to accelerate and simplify the ML service development. Ease.ml \cite{li2018ease} is an AutoML system which automatically select proper models for different ML tasks in a multi-tenant environment. Rafiki \cite{wang2018rafiki} is a ML analytics service system which provides on-line data analytics capability to support database system. KubeFlow is combination of Kubernetes and TensorFlow which provides an end-to-end ML development pipeline. 

\textbf{Resource management for ML workload.} 
ML development relies heavily on the high-performance computational resources, and how to efficiently manage GPUs in a multi-tenant environment is another important research topic. Optimus \cite{peng2018optimus} uses an online profiling model to find relative resource configuration to reduce training time. Gandiva \cite{xiao2018gandiva} analyzes the training characteristics of the application and uses migration, packing and preemption to speed up the overall performance. 
Tiresias \cite{gu2019tiresias} proposes a job-agnostic discretized two-dimensional least-attained service (DLAS) scheduling by combining the time job has run and the number of GPU the job uses.  \HBS{In comparison, this demo complements those existing studies in the following two aspects. First, existing studies limit the cluster scale in dozens or hundreds of GPUs, and AnalySIM does not have this limitation. Our traces are collected from a cluster of thousands of GPUs. Second, AnalySIM has evaluated many core optimizations in each of existing algorithms such as resource allocation policies and scheduling, which are helpful to understand the design options in a cluster with thousands of GPUs.}

\section{System Overview}\label{sec:overview}
This section describes the design of our cluster simulator AnalySIM. We will first talk about our design principles: 1) \textbf{Fidelity.} The simulator should simulate the real cluster with small errors. 2) \textbf{Portability.} It should be flexible enough for various cluster management policies. 3) \textbf{Analyzability.} It is desirable to provide visualized simulation results in an intuitive way for better analysis.

\begin{figure}[b]
    \centering
    \includegraphics[width=3.1in]{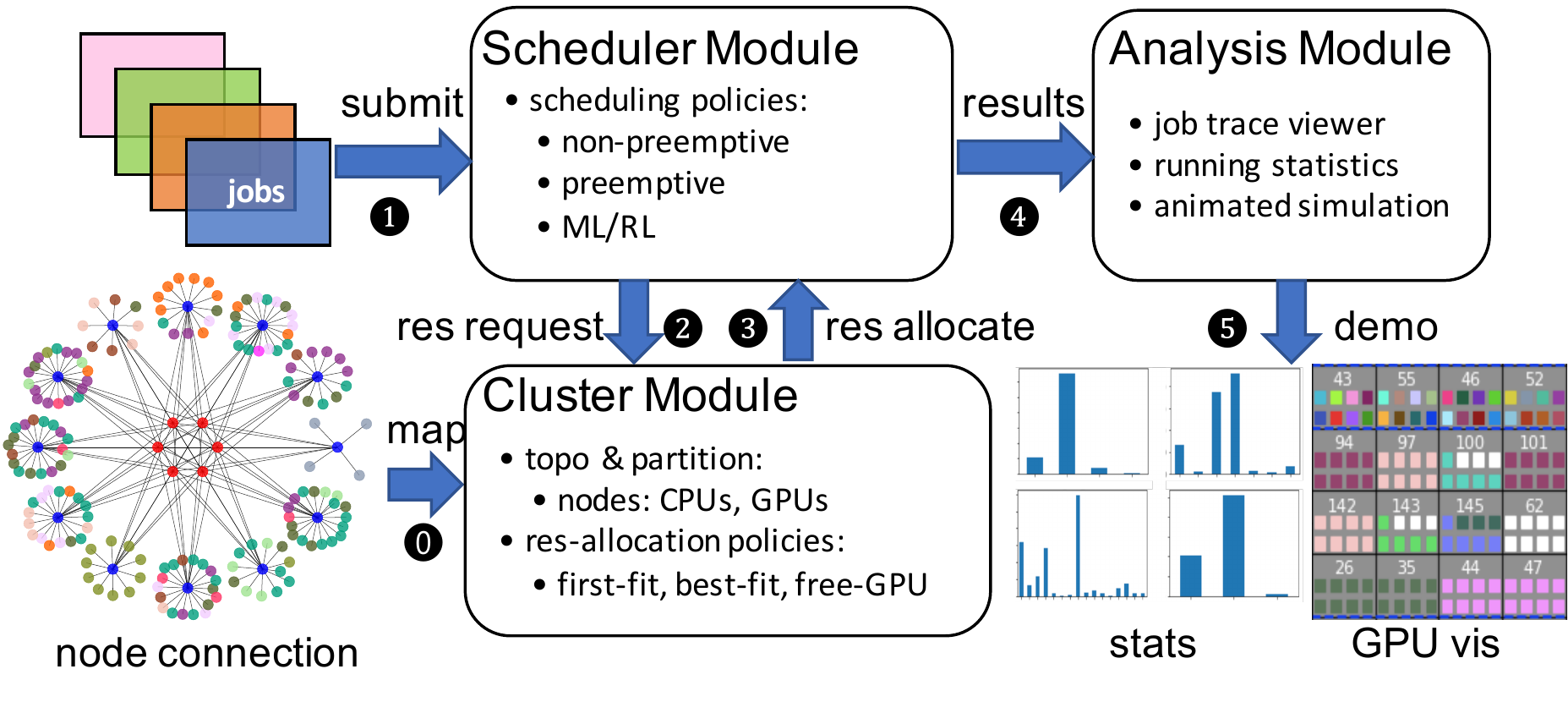}
    \caption{The design of AnalySIM.}
    \label{fig:overview}
\end{figure}

Fig. \ref{fig:overview} shows the general architecture of AnalySIM. Taking the job trace and cluster topology as input, AnalySIM performs the simulation and offers visualization tools for cluster performance and GPU utilization analysis. This simulator is written by Python with Object-Oriented Programming.

\subsection{Cluster Module\label{cluster-module}}
The simulated cluster manages the computational resource of each node in the real cluster. (\texttt{AnalySIM.Cluster} class in Python).
To better characterize the real GPU cluster, we take both physical and logical hierarchies into consideration:

\textbf{Physical topology.} By reading the configurations of cluster manager (e.g., \texttt{topology.conf} in Slurm), we can get the interconnection topology among computational nodes in the cluster. \HBS{We develop programs on sending/receiving data packets of different sizes to calibrate the performance (including latency and bandwidth). Such calibrations are able to apply to different networks such as RDMA and Infiniband. } In order to include the network communication cost, AnalySIM adds the estimated communication overhead for the distributed ML jobs with nodes from different switches. Such overhead is roughly estimated by a penalty ratio obtained from the comparison of data transfer rates among different nodes.

\textbf{Logical partition.} Modern cluster management tools usually \emph{logically} separate the whole computational nodes into different partitions. Such partitioning avoids the resource contention among jobs from different task groups. With the information provided by cluster manager. 
AnalySIM can construct such partitioning on the simulated computational nodes. Each partition has its own resource pool and job queue, which also enables the hybrid partition-level resource allocation and job scheduling policy (e.g., different partitions use different management policies).\par

In our implementation, Cluster class maintains a list of nodes (\texttt{AnalySIM.Node} class), and each node maintains its GPU usage. Besides that, Cluster class also has two maps switches and partitions for representing physical and logical structure of the real cluster.\par

As for computational resource allocation, AnalySIM allows users to select or add different resource allocation policies. Currently, AnalySIM provides three resource allocation policies to serve resource requests in the form of the tuple (\#Nodes, \#GPUsPerNode).
1)
\textbf{First-fit}. The first available node combination searched will be allocated to the pending job.
2)
\textbf{Best-fit}. All available node resource combinations will be sear\-ched, the smallest sufficient one will be allocated to the pending job.
3) 
\textbf{Free-GPU}. 
Cluster manager can use idle GPUs from any nodes to meet the job requirement of the total \#GPUs needed. From all idle GPU combinations, the resource combination with the smallest number of nodes will be allocated to the pending job. In constrast to first-fit and best-fit, free-GPU allocation can be a better policy which requires the co-design of ML front-end framework (e.g., TensorFlow and PyTorch) and back-end resource management tools.
\par

In addition to the resource allocation policies, AnalySIM supports the simulation of dynamic resource migration for the running jobs. That is, once a running job completes and releases its computational resources, cluster manager will examine those running jobs with least required sources and perform migration if the migration can improve the overall node utilization. \HBS{We estimate the PCI-e overhead of job migrations as the data transferred divided by the PCI-e peak bandwidth.} This mechanism can also help to mitigate the computational resource (GPU) fragmentation problem.\par

\subsection{Scheduler Module}
The scheduler module is used to schedule submitted jobs from collected job traces. \HBS{The trace used for simulation is collected with the cluster manager from the production cluster. We use provided APIs, e.g., \texttt{sacct} in SLURM, to get related job information including job id, job name, the cluster partition the job belongs to, the number of requested nodes and GPU, job’s final state and the times when the job is submitted, started and ended.}

The main purpose of AnalySIM is to analyze the cluster scheduling algorithms and to speed up the development of new scheduling algorithms for GPU cluster. To fully support such purpose, AnalySIM provides flexible interfaces for the implementation of various types of scheduling algorithms. By providing  predefined  template  function interfaces for cluster module and scheduler module, we make different types of scheduling and mechanisms able to run on our simulator. We categorize cluster scheduling algorithms into the following three types: \par

\textbf{Non-preemptive scheduling.} Once a submitted job gets required computational resources, this job will not be preempted until completion or failure. The FCFS (First Come First Serve) is one of classic non-preemptive scheduling algorithms. 
Based on the FCFS algorithm, many other non-preemptive scheduling algorithms like SJF (Shortest Job First) and LAS (Least Attained Service) \cite{nuyens2008foreground} can be implemented with some minor modifications. 

\textbf{Preemptive scheduling.} This type of scheduling allows the running jobs to be preempted and to be resumed running later. One typical preemptive scheduling algorithm is Round-Robin  scheduling. 
Based on preemptive interface provided by AnalySIM, other more sophisticated preemptive scheduling algorithms like MLFQ (Multi-Level Feedback Queue) \cite{epema1995analysis} 
can also be easily implemented. For example, we have implemented LAS-MLFQ scheduling as a hybrid of LAS~\cite{nuyens2008foreground} and MLFQ,
which gives each job different quantum units according to the number of GPUs that the job uses. If a job uses 4 GPUs, then its quantum unit is 1/4 of a job using 1 GPUs.

\textbf{Learning based scheduling.} Machine learning or reinforcement learning (RL) nowadays can also be applied to the cluster scheduling algorithms. For example, 
Decima \cite{mao2019learning}  uses RL model to schedule big-data jobs. 
Usually, learning-based algorithms need long history job traces as training data for iterative training, which is impracticable on a real cluster. 
In this case, AnalySIM can be used as a flexible virtual environment, which efficiently gives scheduling feedback (e.g., number of running jobs within a time period \cite{mao2019learning}) to the training agents and further speed up development of learning-based scheduling algorithms.
\par

\begin{figure}[tbhp]
    \centering
    \includegraphics[width=3.0in]{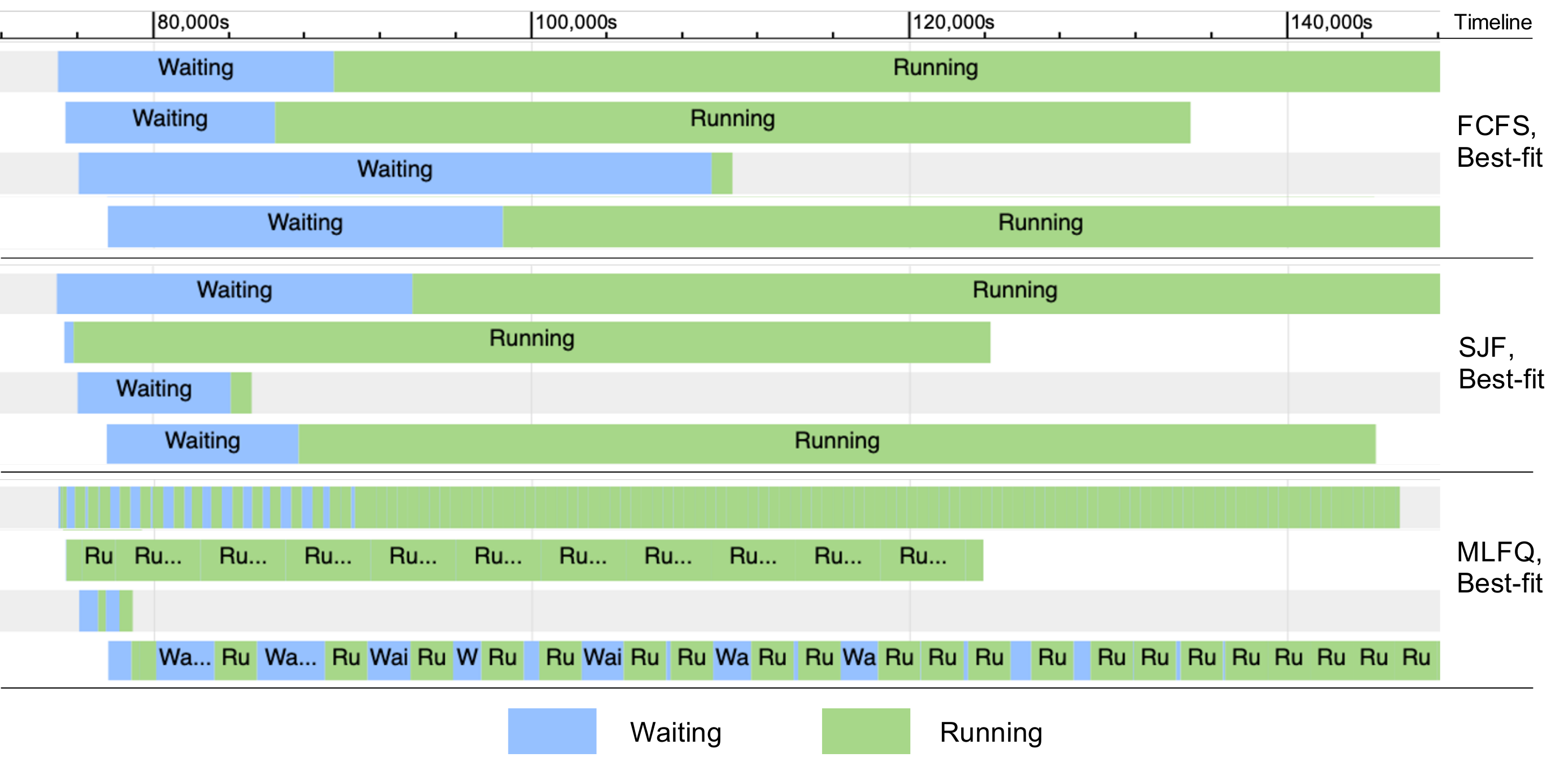}\textbf{}
    \vspace{-6pt}
    \caption{The snapshot of Chrome Trace Viewer for three different cluster scheduling policies. }
    \label{fig-traces}
    \vspace{-15pt}
\end{figure}

\subsection{Analysis Module}
\label{analysis_module}
After getting the scheduling results from scheduler module, the analysis module offers the following functionalities:
1) checking job status and analyzing cluster scheduling efficiency;
2) checking GPU status and analyzing GPU resource utilization. \par
\textbf{Job status analysis.} AnalySIM will generate a simulated job trace containing all job running information. 
By visualizing the trace file, we can directly see how each submitted job is scheduled. Fig. \ref{fig-traces} shows snapshots of Tracing Viewer of different scheduling policies, which gives an intuitive illustration of how different scheduling policies work. Developers can analyze the job scheduling behaviour based on this visualization. 

\begin{figure}[thbp]
  \centering
  \includegraphics[width=3.0in]{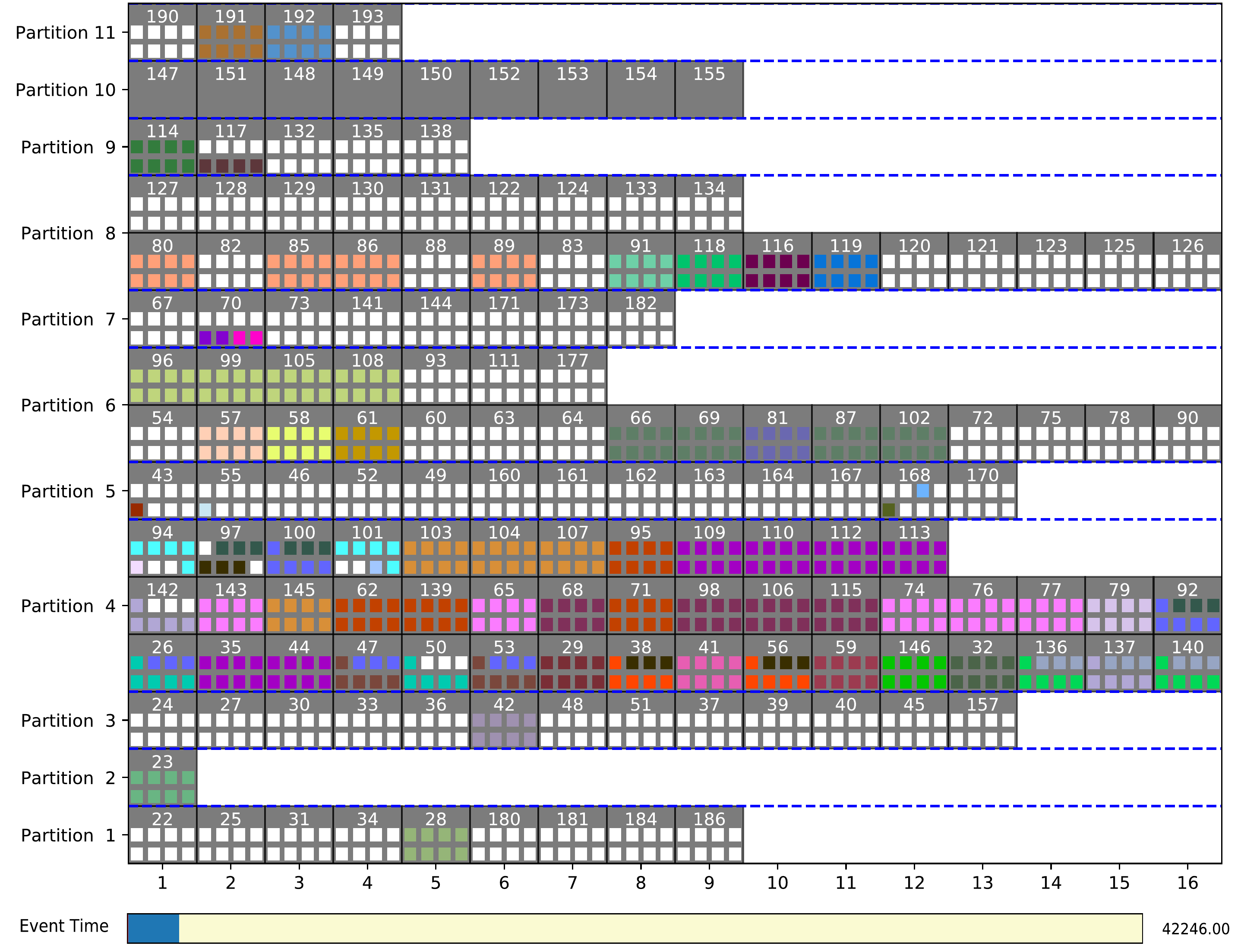}
   \vspace{-5pt}
  \caption{The snapshot of the visualized cluster nodes. Each small square represents a GPU card, different colors on GPUs represent different running jobs, white color represents the idle state. A larger square represents a computational node (with node ID on its top). Nodes are grouped in partitions.}
  \label{demo}
  \vspace{-15pt}
\end{figure}

\textbf{GPU resource utilization and status analysis.} 
Cluster module in AnalySIM also keeps tracing the resource usage of each node.
By utilizing such information, AnalySIM can give a detailed GPU-level visualization together with other cluster status metrics. We implemented these functionalities via Python's Matplotlib package.
Fig. \ref{demo} shows the snapshot of this visualization tool. By moving the time slider (at the bottom), users can study how node utilization changes over time. Through this visualization tool, the bottleneck of a scheduling algorithm can be found in a more straightforward way. For example, adding more nodes to Partition 4 in Fig. \ref{demo} would be helpful to reduce the number of coming pending jobs.\par

\section{Demonstration}
AnalySIM has been used in SenseTime Group Ltd\HBS{, and demonstrated as a viable solution to  understand and evaluate different scheduling policies with the trace from a real production cluster.} In our demonstration, we will use AnalySIM to simulate a real production GPU cluster with 154 nodes and 1160 GPU cards. To run the trace-driven simulation, we use a subset of 20 days' production cluster trace containing 9356 submitted jobs. Almost 40\% of the submitted jobs in this trace require more than 4 GPU cards.


For the resource allocation policy, AnalySIM includes all the scheduling policies described in Section \ref{sec:overview}. For the MLFQ based scheduling, we choose 3 level queues and quantum units for each job queue are 3250, 7200, 18000 seconds, which is adopted from \cite{gu2019tiresias}. As mentioned in \cite{xiao2018gandiva}, the time overhead of deep learning training job stopping and resuming is approximately 8 seconds, we also add 8 seconds as extra overhead for preemption and migration operations in our simulator (nevertheless, this parameter can be configured according to the measurement in the actual cluster). 

\HBS{In the demo, audiences are able to specify different inputs to AnalySIM, such as specifying cluster topology and the number of GPUs, and creating new job traces. Then, they can observe the following demo scenarios in the GUI of the analysis module.  }

To validate the accuracy of our simulation, we study the measured and simulated performance. We find that our simulation has a very high accuracy. The current cluster uses a FIFO-based backfill scheduling policies. The average job completion time (JCT) is 6604.11 seconds in our simulation and 6606.35 seconds in real cluster. This small differences show the high fidelity of our simulation. 

\begin{figure}[t]
  \centering
  \subfigure[\#pending jobs\label{pending_jobs}]{\includegraphics[width=1.6in]{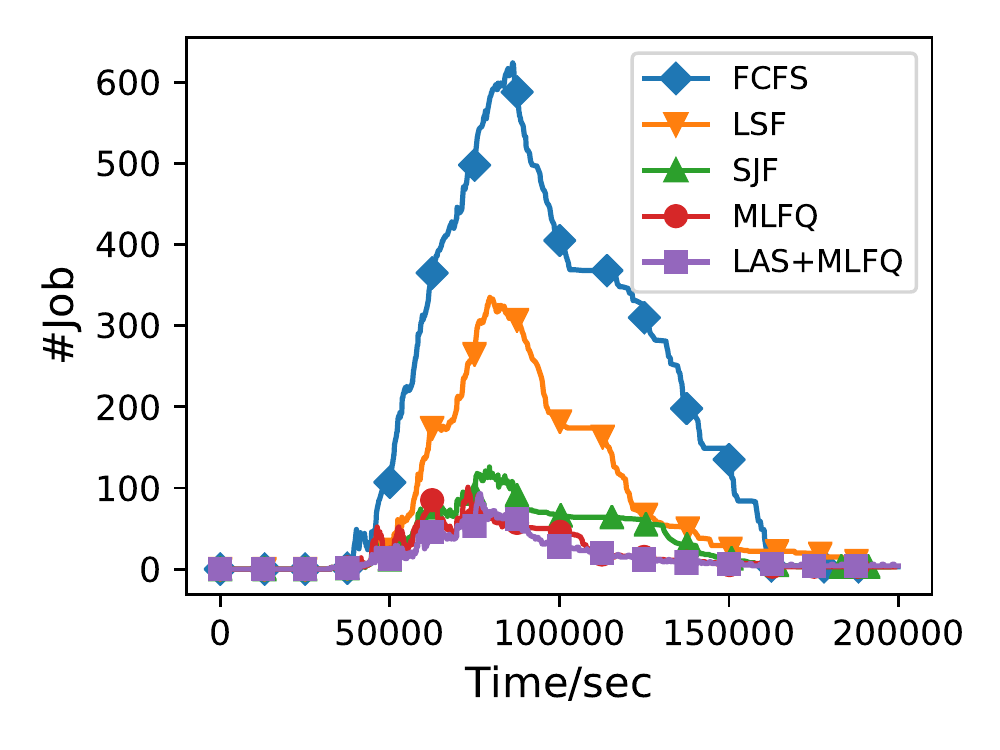}}
  \subfigure[Average JCT(sec)\label{jct}]{\includegraphics[width=1.6in]{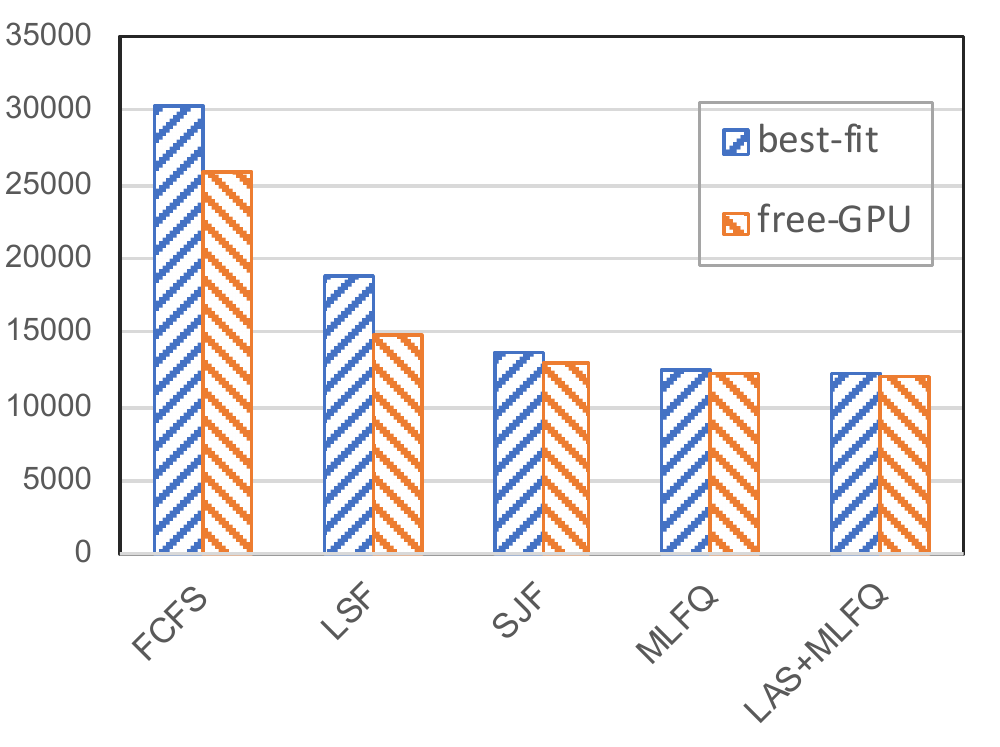}}
  \subfigure[GPU usage.\label{fig:GPU_Ussge} \texttt{+Mig} is for with dynamic migration.]{\includegraphics[width=3.0in]{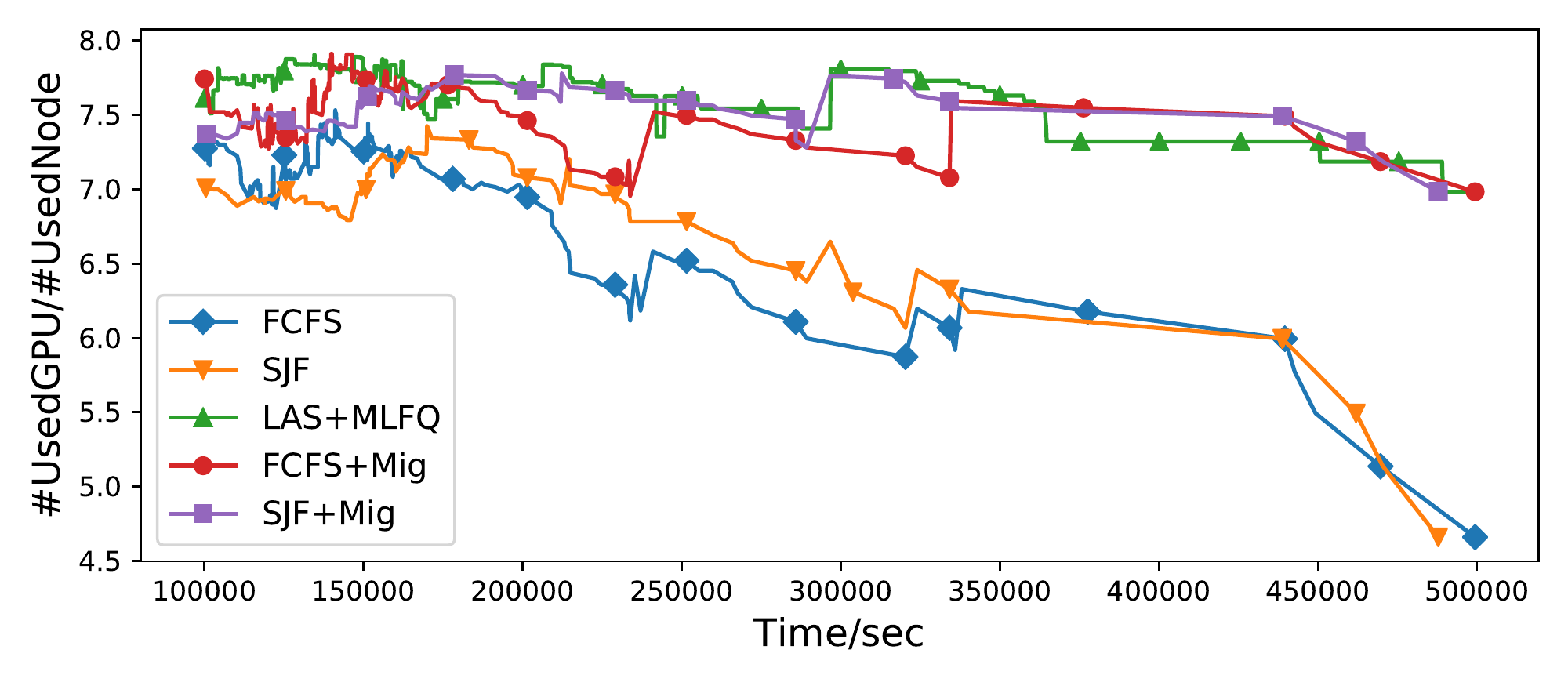}}
  \vspace{-10pt}
  \caption{Running statistics of AnalySIM. Policies in \ref{pending_jobs} and \ref{fig:GPU_Ussge} are under best-fit placement.}
  \label{fig:stats}
  \vspace{-12pt}
\end{figure}

\begin{figure}[b]
  \centering
  \subfigure[Migrations after job completion.\label{mig}]{\includegraphics[width=3.1in]{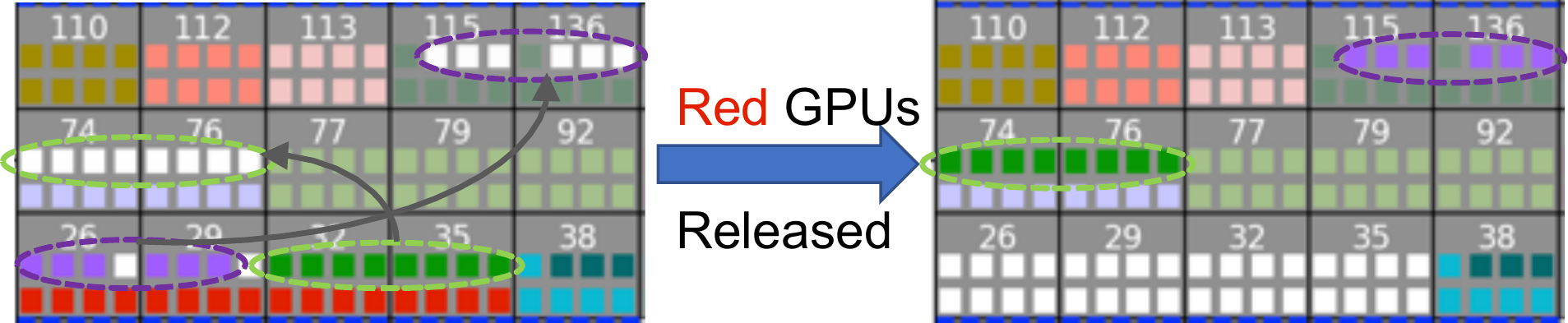}}
  \subfigure[free-GPU placement\label{free-gpu}]{
  \includegraphics[width=3.0in]{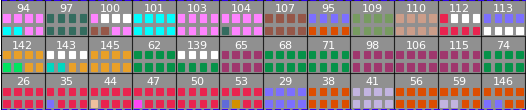}
  }
  \vspace{-10pt}
  \caption{Migration and free-GPU placement. The same figure elements as shown in Fig. \ref{demo}.
  }
  \label{fig:migration_placement}
  \vspace{-12pt}
\end{figure}

\subsection{Scenario 1: Simulation \& Analysis}
We will start by running the AnalySIM simulation with aforementioned scheduling policies to showcase the overall workflow of the AnalySIM, which will let users know how to feed the trace file and cluster configurations into our simulator. By running AnalySIM, users can see the how each node and each GPU is allocated to different jobs along the simulation timeline (i.e. an animated Fig. \ref{demo}). Besides, AnalySIM also plots running statistics (Fig. \ref{fig:stats}) like resource usage rate (\#usedGPU/\#usedNodes), \#pendingJobs and \#runningJobs. By view simulation results like running statistics (Fig. \ref{fig:stats}) and job running timeline (Fig. \ref{fig-traces}), users can easily find the potential problems of a job scheduling policies. As we can see,
there is heavy head-of-line (HOL) blocking for FCFS scheduling, which results in the longer average pending time, while MLFQ solves this problem by doing context switches over running jobs, which mitigate the HOL blocking and ensures the fairness.

\subsection{Scenario 2: Preemption \& Migration}
Fig. \ref{mig} also gives a visual demonstration of how migration works in this simulator. By doing so, a cluster in Fig. \ref{mig} can vacate 4 completely idle nodes for the coming or pending jobs. We can see the improvement brought by preemption mechanism from Fig. \ref{fig:stats}. We use the ratio between the number of GPUs used and the number of nodes used as the measurement of GPU resource fragmentation. Since each node has 8 GPU cards, the ideal GPU utilization ratio is 8. The higher the ratio is, the less resource fragmentation exits. From Fig. \ref{fig:GPU_Ussge}, users can clearly see that preemption and migration mechanisms make the GPU usage at a high rate for a long time, which does mitigate the resource fragmentation problem.
\par

\subsection{Scenario 3: Free-GPU Allocation}
In our simulation, job scheduling with free-GPU placement can always achieve lower average JCT compared with other placements under the same job scheduling policies (Fig. \ref{jct}). Besides, Fig. \ref{free-gpu} shows the GPU status under the free-GPU placement. We can see a job using 48 red GPUs across seven nodes (node 26, 35, 44, 47, 50, 53, 112), though the initial resource requirement of this job is 6 nodes and 8 GPUs per node. If we do not allow such free-GPU allocation, then this large job have to wait pretty long time for 6 wholly unused nodes to be available. In this way, free-GPU can be treated as another complementary way to efficiently deal with fragmented computational resources.

\section{Conclusion}  As machine learning services become emerging, resource management of large-scale GPU clusters becomes a must, which is inefficient, costly and sometimes even risky to production systems. We have demonstrated AnalySIM as an effective platform to research different resource management policies.




\bibliographystyle{ACM-Reference-Format}
\bibliography{sample-base}

\end{document}